\newcommand{\herschel}{\emph{Herschel}}
\newcommand{\hso}{\emph{Herschel}\ Space Observatory}
\newcommand{\etal}{\emph{et al.}}
\newcommand{\msun}{M$_{\odot}$}
\newcommand{\lsun}{L$_{\odot}$}
\newcommand{\rsun}{R$_{\odot}$}
\newcommand{\mum}{$\mu$m}
\newcommand{\eg}{{\it e.g.}}

\newcommand{\rstar}{$R_{*}$}
\newcommand{\mdotenv}{$\dot{M}_{env}$}
\newcommand{\rmaxenv}{$R_{env,max}$}
\newcommand{\rminenv}{$R_{env,min}$}
\newcommand{\thetac}{$\theta_{C}$}
\newcommand{\mdotdisk}{$\dot{M}_{disk}$}
\newcommand{\rmaxdisk}{$R_{disk,max}$}
\newcommand{\rmindisk}{$R_{disk,min}$}
\newcommand{\mdisk}{$M_{disk}$}
\documentclass[letter,traditabstract]{aa} 
\usepackage{graphicx}
\usepackage{color}
\definecolor{Blue}{rgb}{0.0,0.0,1.0}
\definecolor{Red}{rgb}{1.0,0.0,0.0}
\usepackage{txfonts}
\begin{document}
   \title{Predicted Colors and Flux Densities of Protostars in The \herschel\thanks{Herschel is an ESA space observatory with science instruments provided by Principal Investigator consortia. It is open for proposals for observing time from the worldwide astronomical community} PACS and SPIRE Filters}
   \author{Babar Ali\inst{1}
           \and J. J. Tobin\inst{2}
           \and W. J. Fischer\inst{3} 
           \and C. A. Poteet\inst{3}
           \and S. T. Megeath\inst{3}
           \and L. Allen\inst{4}
           \and \\L. Hartmann\inst{2}
           \and N. Calvet\inst{2}
           \and E. Furlan\inst{5}\fnmsep\thanks{Spitzer Fellow}
           \and M. Osorio\inst{6}
          }
   \institute{NHSC/IPAC, California Institute of Technology, 
              Pasadena, CA USA\\
              \email{babar@ipac.caltech.edu}
              \and
              University of Michigan,  Ann Arbor, Michigan USA
              \and
              University of Toledo, Toledo, OH USA
              \and
              NOAO, Tucson, AZ USA
              \and
              JPL, Caltech, Pasadena, CA, USA
              \and
              Instituto de Astrofisica de Andalucia, Granada, Spain
              }
   \date{Received ; Accepted}

  \abstract{
Upcoming surveys with the \hso\ will yield far-IR photometry of large samples of young stellar objects, which will require careful interpretation.  We investigate the color and luminosity diagnostics based on \herschel\ broad-band filters to identify and discern the properties of low-mass protostars.  We compute a grid of 2,016 protostars in various physical conÞgurations, present the expected flux densities and flux density ratios for this grid of protostars, and compare \herschel\ observations of three protostars to the model results.  These provide useful constraints on the range of colors and fluxes of protostar in the \herschel\ filters.  We find that \herschel\ data alone is likely a useful diagnostic of the envelope properties of young stars.}
   \maketitle
%
%________________________________________________________________

\section{Introduction}
A comprehensive theory of star formation is necessary for understanding processes spanning the entire range of cosmic evolution, from the formation and evolution of galaxies to the formation of planets.  Thus, observing the formation of stars is one of the prime science objectives of the \hso\ \cite{herschel}.  With its large (3.5 m) primary mirror, \herschel\ is able to survey a substantial fraction of the current star formation in the Galaxy with unprecedented efficiency and sensitivity.  The result will be an explosion of far-IR and sub-mm data on star formation regions.  \herschel\ is particularly suited for carrying out these surveys because the spectral energy distributions (SEDs) of protostars peaks in the far-IR and sub-mm region.  A large number of guaranteed- and open- time key (legacy) \herschel\ programs are, thus, focused on star formation regions.  These include PACS (photodetector array camera \& spectrometer, \cite{PACS}) and SPIRE (spectral \& imaging reciever, \cite{SPIRE}) imaging surveys of the Gould belt (\cite{andre2010}), and the Herschel Orion protostar survey (HOPS) and DIGIT programs, among others.
\par
These surveys will detect hundreds of thousands of protostars.  For our purposes, protostars are low-mass stars in the earliest stages of their evolution.  They are less than a few~$\times$~10$^5$~yrs old and are still surrounded by an infalling envelope.  The SEDs of these protostars will be complex mixture of their intrinsic emission coupled with reprocessing and emission from their immediate environment.  Radiative transfer models are an important tool in understanding and interpretting the observed flux densities of protostars.  Model SEDs can provide crucial constraints on the most important physical parameters of a protostellar system.  Because protostars can encompass a wide rage of parameter space it is often necessary to compare observations against a large number of different models to derive a reasonable solution and to understand the degeneracies involved.  \cite{robitaille2006} constructed a grid of 20,000 individual models for SED fitting; however, these models used an ism model for the envelope, which is not accurate at far-IR wavelengths.  This is because reprocessing by dust is especially important at \herschel's wavelengths as this reprocessed light forms the bulk of the protostars' SED.   In this contribution, we have generated a grid of 20,160 SEDs from 2,016 individual protostar\footnote{There are 10 different inclination angles for each of 2,016 protostars}  models focused primarily on the parameter space appropriate for young stars, and for the \herschel\ broad-band filters.

\section{The model grid and flux predictions}
\label{SecModel}
\begin{figure}
\centering
\includegraphics[width=8cm]{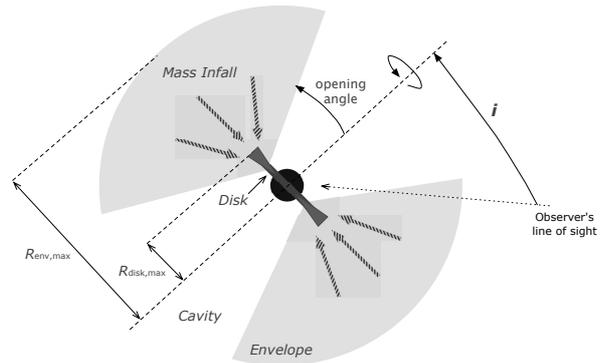}
\caption{Illustration of some of the major parameters used to model the protostar SEDs.}
\label{FigPars}
\end{figure}

We use the Monte Carlo radiative transfer code developed by \cite{whitney2003}, which is adapted to include the dust mixture properties  from Tobin~\etal (2008) to calculate the model SEDs.  For convenience, we briefly summarize the main components here.  The code comprises four physical components: (i) a central object.  (ii) A flared disk.  (iii) A rotating collapsing envelope, and (iv) a bipolar cavity.  These components are illustrated in Fig.~\ref{FigPars} .  The star+disk system emits energy packets ("photons") that are absorbed or scattered by the disk and/or envelope, subsequently determining the emergent SED of the system.  Our adopted dust model uses the methodology described by \cite{dustmodel} with additional optical constants for graphite from \cite{dustgrains}.  The dust grains used here are a mixture of graphite, silicate and water ice and their number density has a power-law distribution with grain size to the power -3.5 and include large (1 micron radius) grains; the implied gas-to-dust ratio is 133.
\par
The code was used to produce a grid of 20,160 model SEDs using 2$\times10^7$\ photons for the Monte Carlo computations, and by using various configurations for the main components listed above.  We restricted the variable parameter set to: the infall rate (\mdotenv), the maximum envelope size (\rmaxenv), the cavity opening angle (\thetac), the cavity shape exponent ($b$), the accretion rate (\mdotdisk),  the size of the disk (\rmaxdisk), and the inclination angle of the system to the observer's line of sight ($i$).   These are the critical parameters that predominantly determine the flux density prediction for \herschel\ PACS and SPIRE bands (56-610~\mum).  The remaining parameters have fixed values that are typical for low-mass protostars (\cite{tobin2008}).  
\par
One important characteristic of the code is that the disk accretion rate is used in the model grid to vary the luminosity of the protostellar systems;  the central star is fixed at 1~\lsun. This is reasonable since we do not know the relative contribution of photospheric and accretion luminosities in protostars. However, the  spectrum of the central object's emission does not largely affect the emergent spectrum of the protostar in the far-IR since nearly all emission is reprocessed.  Thus, dust opacities are a more critical and relevant factor for our analysis.
\par
Table~\ref{TblPars} summarizes the relevant model parameters and their values in our grid.  We list only select parameters (see \cite{tobin2008} for a complete list).  The properties of the central object are perhaps the most crucial of the fixed parameters in that changing the parameters for central object may lead to degeneracies in the observed colors and fluxes.  We discuss this further in Sect.~\ref{SecScaling}.   Other relevant limitations of the model are that dust opacities, which are known to vary across regions, are assumed to be constant, and that only the water ice opacities for the 3.6~\mum\ feature are included in addition to dust.

\subsection{Synthetic Photometry}
The code computes model SEDs for a given spatial extent (aperture) of the system.  We use aperture radius of 20,000 AU.  This collects the total integrated flux from the envelopes which are all smaller than the aperture.  We convolved the resulting SEDs with the filter+detector transmission curves of PACS and SPIRE (kindly provided by their respective instrument teams).   We present the results as flux density values at the central wavelengths of the filters.  The model fluxes are calculated at the distance of Orion, adopted to be 420~pc (\cite{menten2007}).

% Figure of protostar model parameter.

% The table with model parameters.
   \begin{table*}
      \caption[]{Model parameters.}
         \label{TblPars}
         \centering
         \begin{tabular}{lll}
         \hline\hline
         \noalign{\smallskip}
            Parameter & Description &  Value(s)\\
          \noalign{\smallskip}
          \multicolumn{3}{l}{\underline{Central object \& observer:}}\\
         $M_{*}$(\msun)  & Mass of the central object.  & 0.5\\
         \rstar\ (\rsun) & Radius of the central object. & 2.09\\
         $T_{*}({\rm K})$  & Surface temperature of the central object. & 4000\\
         $i$\ (degrees) & {\bf Inclination angle} & 18.2, 31.8, 41.4, 49.5, 56.7, 63.3, 69.5, 75.6, 81.4, 87.2\\
         \noalign{\smallskip}
         \multicolumn{3}{l}{\underline{Envelope Properties:}}\\
         \mdotenv\ ($\times10^{-6}\ $\msun\ yr$^{-1}$)& {\bf Mass infall rate} &  1, 5, 7.5, 10, 25, 50, 100\\
         \rmaxenv\ (AU) & {\bf Envelope outer radius} & 5000, 10000, 15000 \\
         \rminenv\ (\rstar) & Envelope inner radius & 42.75 \\
         $R_C$\ (AU) & Centrifugal radius & Equal to \rmaxdisk \\
         \noalign{\smallskip}
         \multicolumn{3}{l}{\underline{Disk properties:}}\\
         \mdisk\ (\msun)  & Disk mass & 0.05 \\
         \mdotdisk\ ($\times10^{-8}$\ \msun\ yr$^{-1}$) & {\bf Disk accretion rate/Luminosity\tablefootmark{a}} & 1, 10, 100, 1000\\
         \rmaxdisk\ (AU) & {\bf Disk outer radius} & 5, 100, 500\\
         \rmindisk & Disk inner radius &  Set by dust destruction radius.\\
         $\alpha$ & disk radial density exponent & 2.125 \\
         $\beta$ & disk scale height exponent & 1.125 \\
         \noalign{\smallskip}
         \multicolumn{3}{l}{\underline{Cavity properties}}\\
         \thetac\ (degrees) & {\bf Cavity opening angle} & 5, 15, 25, 45\\
         $b$ & {\bf Cavity shape exponent} & 1.5, 2.0 \\
\hline
\end{tabular}
\tablefoot{\\
Parameters in bold are model grid variables.\\
\tablefoottext{a}{See Sect.~\ref{SecModel} for the coupling between \mdotdisk\ and luminosity}
}
\end{table*}

\section{Results and discussion}

\begin{figure}
\centering
\includegraphics[width=8cm]{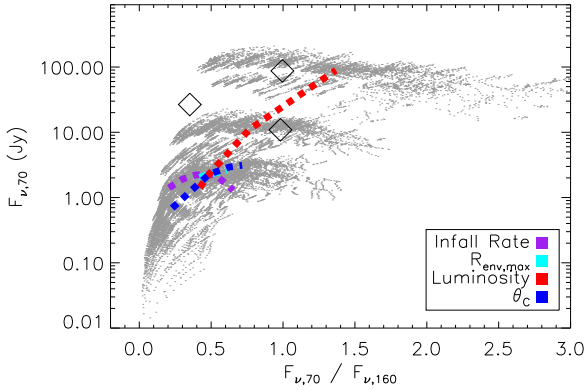}
\caption{Predicted results for the PACS 70~\mum\ filter {\it vs.}\ the ratio of 70 to 160~\mum\ flux density.  The fluxes are reported for a protostar at the distance of Orion.  The shaded region shows the full range in the 20,160 SEDs in the grid.  The color-coded (inset) thick dashed lines illustrate the magnitude and direction of the parameter when all other parameters are held constant.  We do this by choosing typical values (see text) and plotting the variation from only one parameter for that protostar.  The diamond symbols show the locations of the 3 HOPS protostars (see Sect.~\ref{SecHOPS}).}
\label{FigPACS}              
\end{figure}

\begin{figure}
\centering
\includegraphics[width=8cm]{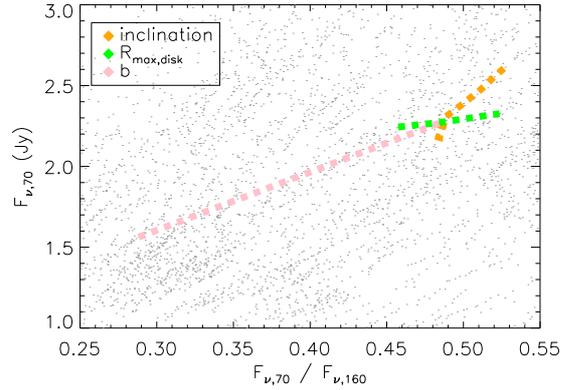}
\caption{Same as Fig.~\ref{FigPACS} but this time for the 3 parameters that appear to produce little or no effect on the observed flux densities.  This Fig. is a magnified portion of the range of values shown in Fig.~\ref{FigPACS}.  The effect of these parameters will not be visible if shown over the full range displayed in Fig.~\ref{FigPACS}.}
\label{FigPACSless}              
\end{figure}

\begin{figure}
\centering
\includegraphics[width=8cm]{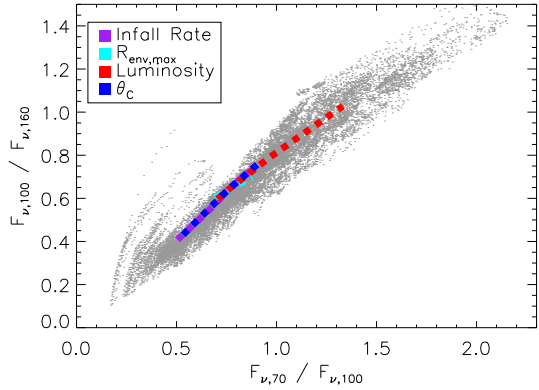}
\caption{The ratio of the PACS flux densities, 70~\mum/100~\mum\ {\it vs.}\ 100~\mum/160~\mum.  The symbols and lines are as shown in Fig.~\ref{FigPACS}.}
\label{FigPACSclrs}              
\end{figure}

\begin{figure}
   \centering
   \includegraphics[width=8cm]{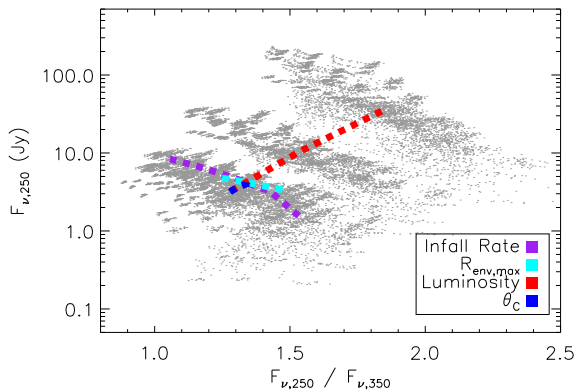}
   \caption{The SPIRE 250~\mum\ flux density {\it vs.} the ratio 250~\mum/350~\mum.  The symbols and lines are as in Fig.~\ref{FigPACS}.}
   \label{FigSPIRE}              
\end{figure}

\begin{figure}
\centering
\includegraphics[width=8cm]{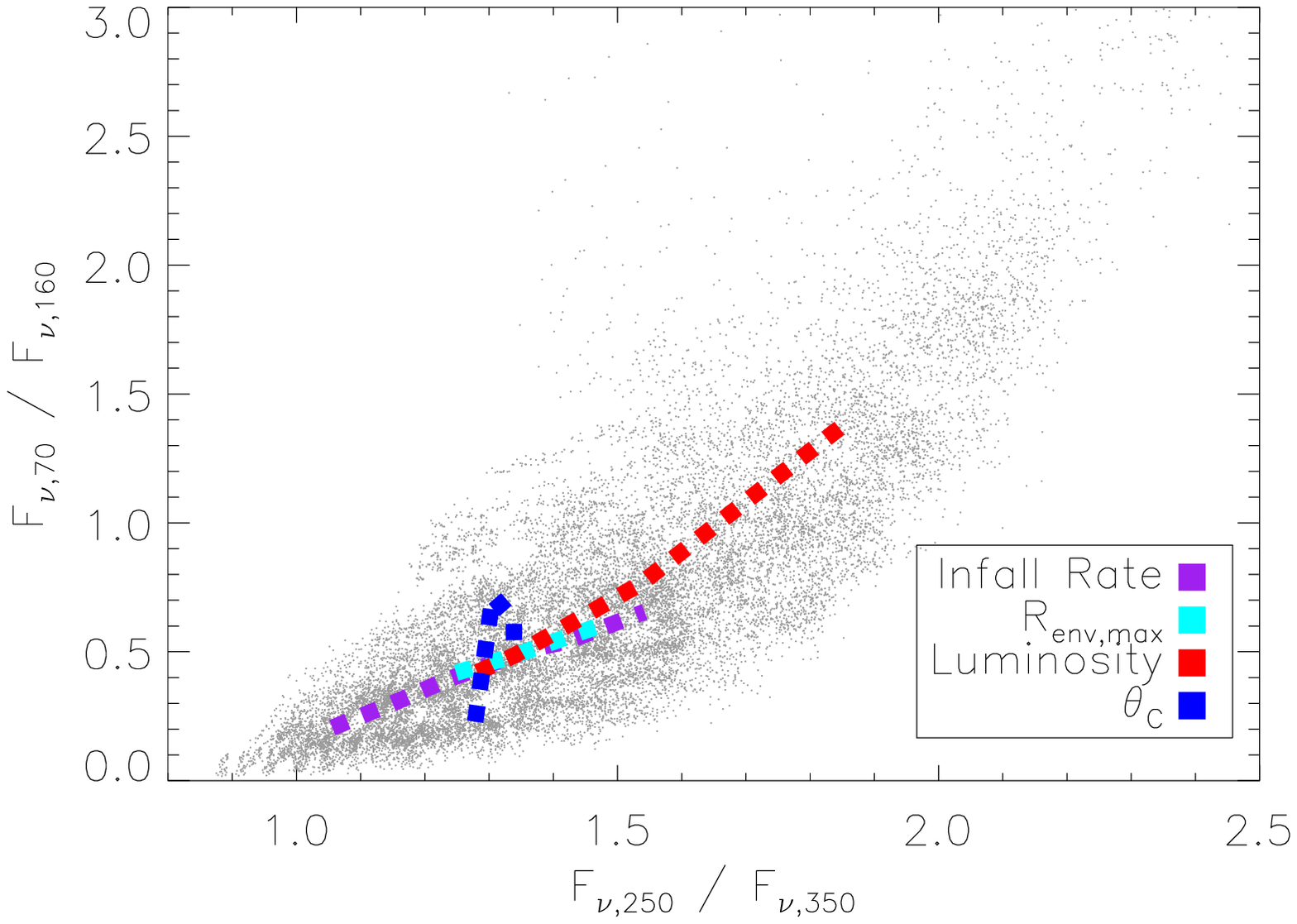}
\caption{The PACS (70\mum/160\mum) ratio {\it vs.}\ the SPIRE (250\mum/350\mum) ratio.}
\label{FigPACSSPIRE}              
\end{figure}

Figures~\ref{FigPACS}$-$\ref{FigPACSSPIRE}\ show the results for the 20,160 model SEDs calculated for our grid for various combinations of flux densities and flux density ratios.  The shaded dots shows the full range of predicted flux densities for the entire set of 2,016 protostar systems as seen with 10 different inclination angles each.   The parameter values used here sample a broad range of physical conditions in the protostars and, thus, provide a reasonable comparative range of expected flux densities and flux density ratios for protostars.
\par
In protostellar systems, model grids, such as the one discussed here, are most accurate and useful when multi-wavelength data are available to provide as many constraints as possible (see \eg\ \cite{hops}).  Nonetheless, PACS and SPIRE sample the peak of the protostar's SED and we expect \herschel\ data alone to provide a relatively good estimate of the envelope properties in particular.   
\par
We further investigate the effect of each of the variable parameters on the observed fluxes.  To illustrate how a particular model parameter affects the flux densities, we first selected a typical protostar system, here defined to have the following values:  \mdotdisk=$10^{-7}$~\msun~yr$^{-1}$, \rmaxdisk=100~AU, \mdotenv=$10^{-5}$~\msun~yr$^{-1}$, \rmaxenv=10,000~AU, \thetac=25~degrees, $b=1.5$, and $i=63.3$~degrees.  Then, we show how individual parameters affect the results as they vary over the range listed in Table~\ref{TblPars}, while holding all other parameters constant.  The results are color-coded (as listed in the legend boxes in the upper left hand part of Figs.~\ref{FigPACS}$-$\ref{FigPACSSPIRE}) and shown as thick dashed lines.  We show and discuss below results from a representative set of correlations investigated from the full grid of model parameter range.  We omit discussion on \mdotdisk, as this parameter and its use in the code is already discussed above (see Sect.~\ref{SecModel}). 

\subsection{Envelope Infall rate}
The mass infall rate strongly affects the far-IR flux and not surprisingly, the predicted flux density and their ratios typically vary by factors $\sim$2\ as a function of  \mdotenv.  This is straight forward as increasing the amount of infall simply increases the amount of cold material and thus the far-IR fluxes.  We expect to be able to constrain the envelope infall rates with relatively good precision with {\emph{Herschel}}\ data alone, although, some knowledge about the distance to the source (or its luminosity) is required, or must be assumed.

\subsection{Envelope size}
The size of the envelope directly affects the amount of the coldest dust present in the protostar.  Thus, we note significantly more variation in the longer wavelength SPIRE filters (see Figs.~\ref{FigSPIRE}$-$\ref{FigPACSSPIRE}) than in the PACS filters.  Increasing the outer radius of the envelope preferentially creates a comparatively redder protostellar system.  However, the affect of the \rmaxenv\ is still appreciably lower than \mdotenv.  That's because the latter more directly controls the envelope density and total amount of emitting material present in the envelope.

\subsection{Cavity opening angle}
The PACS fluxes appears to be most strongly affected by this parameter (see Figs.~\ref{FigPACS} and \ref{FigPACSclrs}).  This is likely because the width of the cavity preferentially carves away the part of the envelope responsible for the 70~\mum\ flux.   By contrast, there is almost no variation as a function of \thetac\ in the SPIRE filters. 

\subsection{Inclination angle}
The optical and near-IR emission from protostellar systems is dominated by the source itself and the scattered light from the circumstellar envelope.  Thus, face-on models are typically brighter and bluer compared to those in which the central star is either wholly or partially obscured by the disk.  At optical and near-IR wavelengths, these objects are quite often extremely faint or undetected even in unbiased photometry surveys due to high obscurataion.  Figures~\ref{FigPACSless} shows that in the PACS filters this effect becomes largely unimportant as the main emission is from the envelope.  We find similar results for the other combinations of flux and flux density diagnostics (not shown).

\subsection{The central object}
\label{SecScaling}
We have concentrated on a single central star; however, the models are applicable to all low mass stars.  The observed fluxes and SEDs simply scale with the mass of the central object as determined by the envelope density profile and the accretion luminosity equations.  When the mass of the central star is changed, the envelope density remains the same but the infall rate changes.  And, the mass accretion rate for a given luminosity scales inversely with the central object mass. Thus, considerable overlap does exist in the observed fluxes for different mass protostars.  This result further demonstrates that PACS and SPIRE are best suited for constraining the envelope and disk properties and provide limited constraints on the central object itself.

\subsection{The HOPS protostars sample}
\label{SecHOPS}
The HOPS program (\cite{hops}) imaged 4 protostars in Orion L~1641 region for which PACS 70 and 160~\mum\ photometry is available for 3 sources.  These stars are shown with the diamond symbol in Fig.~\ref{FigPACS}.  Based on this diagnostic alone, we can conclude that all 3 HOPS sources are much more luminous than our typical protostar of 1~\lsun.  Only 3 sources with two-band photometry are in the HOPS field; nonetheless, the sources are not clustered but spread over a broad range of color and flux density values in Fig.~\ref{FigPACS}, implying that a broad range of values must also exist for the other parameters.  These conclusions are indeed confirmed by detailed model SED fits over the 1-870~\mum\ range for the HOPS sources (\cite{hops}).

\section{Conclusions}
The model grid presented here provides the expected flux densities and flux density ratios for 20,160 low-mass protostellar SEDs, useful for determining the phase space occupied by protostar in the \herschel\ filters.  Disentangling the contributions of the various physical components that make up a protostellar system is observationally challenging and requires multi-wavelength data to properly constrain the properties of the system.  Nonetheless, preliminary results show that the fluxes in the \herschel\ PACS and SPIRE filters are dominated by reprocessed light and emission from the protostellar envelope.  \herschel\ data alone is, therefore, expected to be able to discern envelope properties with some precision provided that a reasonable estimate for the object's distance or its luminosity is available.  The source of the emission (combination of central star and disk accretion) is relatively unimportant at \herschel\ far-IR and sub-mm wavelengths.  Thus, \herschel\ data is unlikely to constrain the mass of the central object, but sampling the peak of the protostars SED allows tight constrains on its total luminosity.

\begin{acknowledgements}
We thank B. Whitney for making her code available publicly.
B. Ali acknowledges support from NASA grant IPAC.ALI-OTKP-1-JPL.000094. 
J. J. Tobin acknowledges support from NASA grant HST-GO-11548.04-A.
\end{acknowledgements}


\begin{thebibliography}{}

\bibitem[Andre \etal\ 2010]{andre2010}Andre, P., \etal\ 2010, A\&AS, this volume.
\bibitem[D'Alessio \etal\ (2001)]{dustmodel}D'Alessio, P., Calvet, N., \& Hartmann, L. 2001, ApJ, 553, 321
\bibitem[Draine \& Lee (1984)]{dustgrains}Drain, B. T. \& Lee, H. M. 1984, ApJ, 285, 89
\bibitem[Fischer \etal\ 2010]{hops} Fischer, W., J. \etal  2010, A\&AS, this volume.
\bibitem[Griffin \etal\ 2010]{SPIRE} Griffin, M.,  \etal\ 2010, A\&AS, this volume.
\bibitem[Menten \etal\ 2007]{menten2007}Menten, K. M., Reid, M. J., Forbrich, J., \& Brunthaler, A. 2007, A\&A, 474, 515
\bibitem[Pilbrat \etal\ (2010)]{herschel} Pilbrat, G., \etal\ 2010, A\&AS, this volume.
\bibitem[Poglitsch \etal\ 2010]{PACS} Poglitsch, A., \etal\ 2010, A\&AS, this volume.
\bibitem[Robitaille \etal\ 2006]{robitaille2006}Robitaille, T. P., Whitney, B. A., Indebetouw, R., Wood, K., \& Denzmore, P. 2006, ApJS, 167, 256Whitney, B. A., Wood, K., Bjorkman, J. E., \& Cohen, M. 2003, ApJ, 598, 
\bibitem[Tobin \etal\ 2008]{tobin2008} Tobin, J., J., Hartmann, L., Calvet, N., \& D'Alessio, P. 2008, ApJ, 679, 1364
\bibitem[Whitney \etal\ (2003)]{whitney2003} Whitney, B. A., Wood, K., Bjorkman, J. E., \& Wolff, M. J. 2003 ApJ, 591, 1049
 
\end{thebibliography}
\end{document}